\documentstyle [12pt,epsfig]{article} 
\makeatletter

\@addtoreset{equation}{section}
\makeatother

\newcommand{\ba}{\begin{eqnarray}}
\newcommand{\ea}{\end{eqnarray}}

\newcommand{\Sc}{{\cal S}}
\newcommand{\Lc}{{\cal L}}

\begin{document}
\newcommand{\BS}{\bigskip}
\newcommand{\SECTION}[1]{\BS{\large\section{\bf #1}}}
\newcommand{\SUBSECTION}[1]{\BS{\large\subsection{\bf #1}}}
\newcommand{\SUBSUBSECTION}[1]{\BS{\large\subsubsection{\bf #1}}}

\begin{titlepage}
\begin{center}
\vspace*{2cm}
{\large \bf Derivation of the Schr\"{o}dinger equation from the Hamilton-Jacobi equation
   in Feynman's path integral formulation of quantum mechanics}  
\vspace*{1.5cm}
\end{center}
\begin{center}
{\bf J.H.Field }
\end{center}
\begin{center}
{ 
D\'{e}partement de Physique Nucl\'{e}aire et Corpusculaire
 Universit\'{e} de Gen\`{e}ve . 24, quai Ernest-Ansermet
 CH-1211 Gen\`{e}ve 4.
}
\newline
\newline
   E-mail: john.field@cern.ch
\end{center}
\vspace*{2cm}
\begin{abstract}
    It is shown how the time-dependent Schr\"{o}dinger equation may be simply derived
    from the dynamical postulate of Feynman's path integral formulation of quantum 
   mechanics and the Hamilton-Jacobi equation of classical mechanics. Schr\"{o}dinger's own
   published derivations of quantum wave equations, the first of which was also based
    on the Hamilton-Jacobi equation, are also reviewed. The derivation of
    the time-dependent equation is based on an {\it a priori} assumption equivalent to
    Feynman's dynamical postulate. De Broglie's concepts of `matter waves' and their
    phase and group velocities are also critically discussed. 

 \par \underline{PACS 03.30.+p}

\vspace*{1cm}
\end{abstract}
\end{titlepage}  
 
\SECTION{\bf{Introduction}}

    In the years 1925-1926 modern quantum mechanics was discovered by two separate routes: the matrix
   mechanics of Heisenberg, Born and Jordan~\cite{HeisQM,BJ,BHJ} and the wave mechanics of de Broglie and 
  Schr\"{o}dinger~\cite{SchrodWM1,SchrodWM2,SchrodWM3,SchrodWM5}. 
  The two formalisms were soon shown to be mathematically equivalent by Schr\"{o}dinger~\cite{SchrodWM3},
  Dirac~\cite{DiracTFQM} and Eckart~\cite{Eckart}. 
   Because of the facility of its application to physical problems (both idealised and in relation to experiments)
   via solutions of the  Schr\"{o}dinger equation, text book teaching of quantum mechanics is largely 
  based on the Schr\"{o}dinger formulation in association with consideration of an abstract Hilbert space 
    of quantum mechanical state vectors. 
   \par There exists however a third, independent, formulation of quantum mechanics the conceptual foundations
   of which were laid down in Dirac's 1933 paper `The Lagrangian in Quantum Mechanics'~\cite{DiracLQM}
   which was the basis for Feynman's
  later path-integral formulation~\cite{FeynRMP}. This approach has two great advantages as compared to the earlier
  matrix mechanics and wave mechanics ones: (i) An axiomatic formulation comparable to that 
     provided by Newton's Laws for classical mechanics (ii) Immediate introduction of the key {\it physical}
       concept ---quantum-mechanical superposition---that differentiates quantum from classical mechanics.
      This is of particular importance for the teaching of the subject since, in the conventional appproach,
       it is often difficult to distinguish physical principle from manipulation of the abstract
        mathematical formalism. For example, there is no conceptual difference between seeking solutions of the
      quantum Schr\"{o}dinger equation, to seeking those of any partial differential equation
      of classical physics.
        The relation between projection operations in an abstract Hilbert space and measurements in an
        actual quantum experiment is far from clear.
    \par It is interesting to note that Dirac's work was preceded by similar considerations in a paper written by de
   Broglie in 1923~\cite{dBPM} (cited by neither Dirac nor Feynman) 
  in which the outline of a relativistic electrodynamic theory incorporating photons
  (termed light quanta) obeying the laws of relativistic kinematics was proposed
   \footnote{This paper, predating both matrix mechanics and wave mechanics, actually contains all the
   essential conceptual elements of Feynman's formulation of Ref.~\cite{FeynRMP}
    ---interfering amplitudes (called `waves') that give the probability of observing
    a photon (or any other particle) around a given space-time point. Even the 
   Lorentz-invariant phase of the path amplitude of a free particle (see Eq.~(6.2) below)
   was correctly given, in an added note, together its connection to Hamilton's Principle, as later stated by Dirac.}.
   \par Unfortunately, knowledge of the path integral formulation  has yet to penetrate
    in any significant way into text books on, and the teaching of, quantum mechanics. In one widely used
     book~\cite{Messiah} the name `Feynman' does not even appear in the index! In another, more recent, one~\cite{CCT}
    the postulates
     (I and II of Section 3 below) of Feynman's formulation are stated but no applications are described.
     Notable exceptions are the introductory text book of L\'{e}vy-Leblond and Balibar~\cite{LLB} which is entirely
     based on Feynman's approach, and contains many applications to experiment, as well as the work of
      Taylor~\cite{EFT1,EFT2} that has particularly stressed the unified view of classical and quantum mechanics
      provided by the approach and its mathematical simplicity. In fact, as already pointed out by Dirac~\cite{DiracLQM}, 
     the formalism provides a 
       {\it derivation} of the Principle of Least Action (and hence all of classical mechanics, via the Lagrange
       Equations) as the $h \rightarrow 0$ limit of quantum mechanics. Good introductions to Feynman's axiomatic
       formulation are to be found in Sections 2-4 of Ref.~\cite{FeynRMP}, in Chapter 3 of Volume III of the
       `Feynman Lectures in Physics'~\cite{FLPIII} as well as in the first chapter of the Feynman and Hibbs book~\cite{FH}.\footnote{Feynman's original path-integral paper~\cite{FeynRMP} is included in a recently
          published book~\cite{FeynThesis} which also contains Feynman's PhD thesis on the same subject, as well
          as an introductory preface by L. M. Brown that sets the work in its historical context, as seen 
         from a near-contemporary viewpoint.} A clear presentation of Hamilton-Jacobi theory can be found 
       in Goldstein's `Classical Mechanics'~\cite{Goldst1}. The interested reader who is unfamiliar 
        with the path-integral formulation of quantum mechanics  or Hamilton-Jacobi theory 
     is advised to consult the above sources before reading the present paper.   
     \par Actually, the most detailed working-out of the concepts of the space-time approach to quantum
     mechanics are to be found, not in the research or pedagogical literature, but in the popular
     book `QED the Strange Story of Light and Matter'~\cite{FeynQED}, that Feynman completed shortly before his death.
    In this book many physical effects  ---rectilinear propagation, reflection, refraction and diffraction
     of light, which are considered in standard text books to be examples of applications of
   the `classical wave theory of light' \footnote{See, for example, Born and Wolf~\cite{BW} and, particularly, Jenkins
    and White~\cite{JW} in which the `phase arrows' of Feynman's popular book~\cite{FeynQED} are analysed in full
    mathematical detail.} are shown to be, in fact, pure applications of the space-time
    formulation of quantum mechanics. The detailed mathematical description of many of the
      examples presented, in a qualitative way, in Feynman's book can be found in a recent paper
    by the present author~\cite{JHFAP}. 
    \par The  motivation of the present paper is to improve the general knowledge of the path integral
     formulation of quantum mechanics, and illustrate its importance, by exploring the connections of this
    approach with the more 
     conventional one based on the Schr\"{o}dinger wavefunction. In particular, relations are pointed
    out, in the context of the path integral formalism, between the Hamiltonian description of a
      system in classical mechanics, and the quantum 
    mechanical description of the same system by Schr\"{o}dinger's equation.    
   \par The most evident application of Feynman's formulation is to experiments where path 
     amplitudes of free particles enter essentially into the description of the experiment considered:
      for example, refraction, diffraction and interference experiments in physical optics as discussed
     in Feynman's popular book,
     or `particle oscillations', involving, either neutral bound states containing heavy quarks,
     or neutrinos. Examples of such applications are given in Ref.~\cite{JHFAP} where it is
     also shown that the classical wave theory of light, based on solutions of the Helmholtz 
     equation, is actually a necessary consequence of the path integral formulation of
    quantum mechanics, as it applies to photons interacting with matter. 
    \par The utility of the path integral formulation to the problem, the resolution of which
     gave birth to modern quantum mechanics ---atomic structure--- is less evident. Feynman
     derived, in Ref.~\cite{FeynRMP}, the time-dependent Schr\"{o}dinger equation from the path integral
     postulates. However as will be demonstrated in the present paper, {\it the latter 
     were actually
    implicit in  Schr\"{o}dinger's own original derivation of this equation.}
     \par The action function, $S$, that appears in the fundamental dynamical postulate of the 
        path integral formulation (Eq.~(3.1) below) is actually Hamilton's principle function,
          which is the solution of the Hamilton-Jacobi partial differential equation for the
       system under consideration. The related Hamilton's characteristic function and the corresponding
     Hamilton-Jacobi equation are, as will be seen below, by a devious route, the basis for
    Schr\"{o}dinger's first published derivation of the time-independent Schr\"{o}dinger equation,
    from which the correct energy levels for the hydrogen atom are obtained
    \par The main aims of the present paper are, firstly, to show how the 
    time-dependent Schr\"{o}dinger equation follows directly from the Hamilton-Jacobi
    equation and the dynamical postulate of the path integral formulation and, secondly, to
    review  Schr\"{o}dinger's own derivations of both the time-independent and time-dependent
     quantum wave equations. In order to do this in a comprehensible manner some related topics
     are briefly described in the preceding sections 5, 6 and 7.
    \par The structure of the paper is as follows: In the following section, Hamilton's principle function
     and the Hamilton-Jacobi equation for a classical mechanical system are defined. In Section 3 the 
     postulates of Feynman's path amplitude formulation of quantum mechanics are stated. Section 4
      contains the derivation of the time-dependent Schr\"{o}dinger equation. The Lorentz-invariant phase
     of the path amplitude for a free particle is derived from the transformation equations for
      Hamilton's principle function in Section 5. In Section 6 the de Broglie and Planck-Einstein 
      relations are are obtained by mathematical subsitution in the formula for the phase of the
       path amplitude of a free particle. Also discussed in this section are de Broglie's original
     derivation of the wavelength formula, Hamilton's 1834 analogy between classical mechanics and
     geometrical optics, particle velocity, group velocity, phase velocity and `wave-particle duality'.
      In Section 7, dynamical operators associated with the path amplitude of a free particle
      are introduced together with their eigenvalues. The same operators are shown to respect
      canonical commutation relations and to enable derivation, from the appropriate Hamiltonian
      function, of the time-independent Schr\"{o}dinger equation for the hydrogen atom. Dirac's
     original derivation of the commutation relations, by consideration of their relation to
    the Poisson brackets of classical mechanics, is also discussed.
       Section 8 describes Schr\"{o}dinger's various derivations of the quantum wave equation
      published in 1926. The concluding Section 9 contains a summary of the previous
        sections and a  brief discussion of connections to related work.
      \par The material presented in the present paper is of particular interest to university teachers of
        advanced quantum mechanics courses, at both undergraduate and graduate level, and their students.
        Teachers of quantum mechanics, at all levels, may also benefit from a reading of the paper. Certain topics
       presented in Sections 6 and 7 as well as the whole of Section 8 may be of interest to teachers of, and
       researchers on, the history of science.

  \SECTION{\bf{The Hamilton-Jacobi equation and Hamilton's principle function}}
    A contact, or canonical, transformation applied to the spatial coordinates, $q_i$, and canonical momenta, $p_i$,
    ($i =1,2,...,n$) of an $n-$dimensional classical mechanical system, is one which leaves invariant the form of the Hamilton
   equations:
   \begin{equation}
       \frac{d q_i}{d t} =\frac{\partial H}{\partial p_i},~~~~~~~\frac{d p_i}{d t} = -\frac{\partial H}{\partial q_i}.
   \end{equation}
    The canonically transformed coordinates, $Q_i$, and momenta, $P_i$, then satisfy the Hamilton
   equations: 
      \begin{equation}
       \frac{d Q_i}{d t} =\frac{\partial K}{\partial P_i},~~~~~~~\frac{d P_i}{d t} = -\frac{\partial K}{\partial Q_i}
   \end{equation}
    where the transformed Hamiltonian, $K$, is related to the original one, $H$, by~\cite{Goldst1}:
       \begin{equation}
        K(Q_i,P_i,t) =  H(q_i,p_i,t)+\frac{\partial F}{\partial t}
        \end{equation}
     and, $F$, the generating function of the canonical transformation, may have any of the following
     functional dependences: $F(q_i,Q_i,t)$, $F(q_i,P_i,t)$, $F(p_i,Q_i,t)$ and $F(p_i,P_i,t)$. {\it Hamilton's principle
     function}: $S\equiv F(q_i,P_i,t)$, is the generating function for which the transformed Hamiltonian $K$ vanishes, so
     that:
         \begin{equation}
        H(q_i,p_i,t)+\frac{\partial S}{\partial t} = 0.
        \end{equation}
      The transformation equations relating coordinates and momenta are, in this case~\cite{Goldst1}:
      \begin{equation}
       p_i =\frac{\partial S}{\partial q_i},~~~~~~~ Q_i = \frac{\partial S}{\partial P_i}.
   \end{equation}
       Substituting the first of these equations in (2.4) gives the {\it Hamilton-Jacobi} (H-J){\it equation}:
          \begin{equation}
        H(q_i,\frac{\partial S}{\partial q_i},t)+\frac{\partial S}{\partial t} = 0.
        \end{equation}
       For a single non-relativistic particle of Newtonian mass, $m$, moving in a time-independent potential, $V$, the  H-J equation
       is:
      \begin{equation} 
       \frac{1}{2m}\left[\left(\frac{\partial S}{\partial x}\right)^2+
      \left(\frac{\partial S}{\partial y}\right)^2+\left(\frac{\partial S}{\partial z}\right)^2\right]
      +V + \frac{\partial S}{\partial t} = 0.
       \end{equation}
        This equation is the basis of the derivation of the time-dependent Schr\"{o}dinger equation in Section 4 below.
       The H-J partial differential equation (2.6)  may be solved for the function $S$. This solution gives immediately 
       the equations of motion of the system described by $H$ when $Q_i$ and $P_i$ are
       identified with initial values
       of the coordinates and momenta. The $n$ equations:
      \begin{equation}
      Q_i = \frac{\partial S(q_i,P_i,t)}{\partial P_i}
        \end{equation}
         are solved for the $q_i$ to yield the solution of the dynamical problem:
      \begin{equation}
        q_i = q(Q_i,P_i,t).
      \end{equation}
       The solution for the $p_i$ is then given by the first of Eqs.~(2.5) as:
   \begin{equation}
        p_i = \left.\frac{\partial S(q_i,P_i,t)}{\partial q_i}\right|_{q_i = q(Q_i,P_i,t)}.
       \end{equation} 
       \par Since the $P_i$ are constants, fixed by initial conditions, $S$ is a function only of the $q_i$ and $t$
        so that
     \begin{equation}
     dS = \sum_{i=1}^{n}\frac{\partial S}{\partial q_i}dq_i +\frac{\partial S}{\partial t}dt.
      \end{equation} 
       Making use of (2.4) and the first of Eqs.~(2.5):
      \begin{equation}       
 \frac{dS}{dt} = \sum_{i=1}^{n} p_i\frac{dq_i}{d t} - H = L
      \end{equation}
    where $L(q_i, dq_i/dt,t)$ is the Lagrangian function for the dynamical system. Integrating (2.12)
     identifies $S$ as the indefinite time integral of the Lagrangian:
       \begin{equation}   
        S = \int L dt~~+~~{\rm constant}.
        \end{equation}
       As will be discussed in the following section, in virtue of this equation, Hamilton's principle function $S$
      constitutes the mathematical basis of the Feynman path integral formulation of quantum mechanics.
     \SECTION{\bf{Feynman's path integral formulation of quantum mechanics}} 
      Feynman's path integral formulation of quantum mechanics is based on the following two
      postulates~\cite{FeynRMP}:
      \begin{itemize}
      \item[I] If an ideal measurement is performed to determine whether a particle has a path lying
        in a region of spacetime, the probability that the result will be affirmative is the absolute
        square of a sum of complex contributions, one from each path in the region.
       \item[II] The paths contribute equally in magnitude but the phase of their contribution is the
        classical action (in units of $\hbar$) i.e. the time integral of the Lagrangian taken along the
        path.
        \end{itemize}
        The postulate I is a statement of the superposition property of the amplitudes for different
        paths combined with Born's probabilistic interpretation~\cite{BornPI} of them. Postulate II
        states that the 
        amplitude for a given path is some real function, $A$, times $\exp\{(i/\hbar)S_{BA}([\vec{x}(t)])\}$ where
       \begin{equation}
          S_{BA}([\vec{x}(t)]) \equiv \int_{t_A}^{t_B}L([\vec{x}(t)],t)dt.
       \end{equation}             
        Here $L$ is the classical Lagrangian of the system considered and $[\vec{x}(t)]$ corresponds to 
        a particular space time path between specific (fixed) times $t_A$ and  $t_B$. It can be seen 
        from Eq.~(2.13) that $S$ is actually Hamilton's principle function for the system considered.
        The constant in (2.13) cancels in the definite integral on the right side of (3.1).
         As discussed by Feynman~\cite{FeynRMP,FeynThesis2} $S_{BA}$ is a {\it functional} of the path
         $[\vec{x}(t)]$, since the value of $S_{BA}$ given by (3.1) is different for different paths.
          A particular path  $[\vec{x}(t)^{(j)}]$ can be specified, to any desired precision, by an array
         of space-time coordinates. Considering one spatial dimension:
        \begin{eqnarray}
         &[x(t)^{(j)}] :&~ x_A^{(j)}, t_A^{(j)};~x_1^{(j)}, t_1^{(j)};~x_2^{(j)}, t_2^{(j)};.~.~.
          ~x_n^{(j)}, t_n^{(j)};.~.~.~x_B^{(j)}, t_B^{(j)} \nonumber \\          
            &~&~~~~~~~~t_A^{(j)}<t_1^{(j)}<~t_2^{(j)}.~.~.<~t_n^{(j)}.~.~.~<~t_B^{(j)} \nonumber
         \end{eqnarray}
        Notice that the velocity argument, $\dot{x}(t) \equiv d x(t)/dt$, of the Lagrangian is implicit
        in the specification of the path  $[x(t)^{(j)}]$:
         \begin{equation} 
         \dot{x}^{(j)}(t) = {\rm Lim}(t_{n+1} \rightarrow t_{n-1})~\frac{{x}^{(j)}_{n+1}-{x}^{(j)}_{n-1}}
               {{t}^{(j)}_{n+1}-{t}^{(j)}_{n-1}}. 
      \end{equation}  
       \par Postulate II defines the space-time propagator, Green function, or kernel which gives 
        the amplitude that a particle system with initial space-time coordinate $\vec{x}_A,t_A$ is found at later
        time $t_B$ at  $\vec{x}_B$, having followed a particular path $[\vec{x}^{(j)}(t)]$:
       \begin{eqnarray}
        K_{BA}^{(j)} &\equiv&  K^{(j)}(\vec{x}_B,t_B;\vec{x}_A,t_A) =  A^{(j)}(\vec{x}_B,t_B;\vec{x}_A,t_A)
       \exp \left\{\frac{i}{\hbar}S_{BA}([\vec{x}^{(j)}])\right\}   \nonumber \\
            &\equiv&  A_{BA}^{(j)}\exp \left\{\frac{i}{\hbar}
          S_{BA}^{(j)}\right\} 
         \end{eqnarray}
            where
            \begin{equation} 
            S_{BA}([\vec{x}^{(j)}]) \equiv  \int_{t_A}^{t_B}L([\vec{x}^{(j)}(t)],t)dt. 
            \end{equation}
         The exponential dependence of the propagator on $S_{BA}$ and the definition of the latter in (3.1)
             as the definite time integral of the Lagrangian implies that the propagator has a factorisation
          property:
         \begin{equation}
          K_{BC}K_{CA} = A_{BC}A_{CA}\exp \left\{\frac{i}{\hbar}
          (S_{BC}+S_{CA})\right\} =  A_{BA}\exp \left\{\frac{i}{\hbar}S_{BA}\right\} =  K_{BA}
          \end{equation}
           where $A_{BA} \equiv A_{BC}A_{CA}$. This equation is a manifestation of the property
          of {\it sequential factorisation} of path amplitudes~\cite{FeynRMP,LLB,JHFAP}.
         \par  The total amplitude for finding a system, originally at $\vec{x}_A,t_A$, at $\vec{x}_B$ at $t_B$ 
         is then, according to the postulate I, given by the sum over all allowed paths:
          \begin{equation} 
     K_{BA} \equiv   K(\vec{x}_B,t_B;\vec{x}_A,t_A) =\sum_j  K_{BA}^{(j)} =
           \sum_j  A_{BA}^{(j)}\exp \left\{\frac{i}{\hbar} S_{BA}^{(j)}\right\}.
       \end{equation}
   \par In the following section the propagator, $K$, of a particle moving in a time-independent
     potential, $V$, is related to the conventional quantum-mechanical wavefunction
     of such a particle. It is seen that the fact that the action functional $S$ in (3.3) is just the
     Hamilton's principle function , which is a solution of the H-J equation (2.7), requires the wavefunction
    to satisfy the time-dependent Schr\"{o}dinger equation.
    
  \SECTION{\bf{Derivation from the H-J equation of the time-dependent Schr\"{o}dinger equation}}
      To make the connection between a quantum wavefunction, $\psi$, and a single path amplitude, the
     path labels $j$, $A$ and $B$
      are dropped, the space-time point $\vec{x}_B$, $t_B$ is denoted simply as $\vec{x}$, $t$, and the
      functional dependence of $K$ and $A$ on these space-time coordinates is considered. 
      The wavefunction is then {\it defined} as:
       \begin{equation} 
      \psi(\vec{x},t) \equiv \frac{K}{A} = \exp\left[\frac{i}{\hbar}S(\vec{x},t)\right].
      \end{equation}
       It will be seen, in the following, that when the function $\psi$ is defined in this manner it
       has all the properties of the well-known wavefunction of wave mechanics. In particular, it is a
        solution of the Schr\"{o}dinger equation.  As discussed in Section 8 below, Schr\"{o}dinger used a very 
     similar equation relating $\psi$ and $S$ when the former
     was introduced, for the first time, in Ref.~\cite{SchrodWM1}. Only the all-important factor $i = \sqrt{-1}$
    was missing! 
     \par Inverting (4.1) gives
   \begin{equation} 
            S = -i\hbar \ln \psi
  \end{equation}
      and, with $\vec{x} \equiv (x,y,z)$: 
 \begin{equation} 
 \frac{\partial S}{\partial x} = -\frac{i \hbar}{\psi} \frac{\partial \psi}{\partial x},~~~
 \frac{\partial S}{\partial y} = -\frac{i \hbar}{\psi} \frac{\partial \psi}{\partial y},~~~
\frac{\partial S}{\partial z} = -\frac{i \hbar}{\psi} \frac{\partial \psi}{\partial z},~~~
\frac{\partial S}{\partial t} = -\frac{i \hbar}{\psi} \frac{\partial \psi}{\partial t}.
  \end{equation}
   Transposing the first of Eqs~(4.3), differentiating w.r.t. $x$ a second time:
   \begin{equation} 
     \frac{\partial^2 \psi}{\partial x^2} = \frac{i}{\hbar}\frac{\partial \psi}{\partial x}
        \frac{\partial S}{\partial x}+ \frac{i}{\hbar}\psi \frac{\partial^2 S}{\partial x^2}.
 \end{equation}
   Differentiating, w.r.t. $x$, the first of the transformation equations (2.5) gives:
     \begin{equation} 
     \frac{\partial^2 S}{\partial x^2} = \frac{\partial p_x}{\partial x}= m\frac{d ~}{d t}
     \left(\frac{\partial x}{\partial x}\right) = 0
       \end{equation}
     so that substituting for $\partial \psi/\partial x$ in (4.4) from (4.3), using (4.5), and transposing:
   \begin{equation} 
    \left(\frac{\partial S}{\partial x}\right)^2 = -\frac{\hbar^2}{\psi}\frac{\partial^2 \psi}{\partial x^2}. 
  \end{equation}
    Substituting (4.6) and the corresponding equations for the $y$ and $z$ coordinates as well as the
     time derivative equation in (4.3) in the H-J equation, (3.7), for a non-relativistic particle moving in a time-independent
    potential:
      \begin{equation} 
       \frac{1}{2m}\left[\left(\frac{\partial S}{\partial x}\right)^2+
      \left(\frac{\partial S}{\partial y}\right)^2+\left(\frac{\partial S}{\partial z}\right)^2\right]
      +V + \frac{\partial S}{\partial t} = 0
 \end{equation}
  then yields the time-dependent Schr\"{o}dinger equation:
    \begin{equation} 
     -\frac{\hbar^2}{2m}\nabla^2\psi +V\psi -i\hbar\frac{\partial \psi}{\partial t} = 0
     \end{equation} 
       where
   \[\nabla^2 \equiv \frac{\partial^2 ~}{\partial x^2} +\frac{\partial^2 ~}{\partial y^2}
           + \frac{\partial^2 ~}{\partial z^2}. \]
   Such a brief and elegant derivation of the Schr\"{o}dinger equation directly from the H-J equation
    was mentioned as a possibility by Feynman in his seminal path integral paper~\cite{FeynRMP1}
    but never actually given by him.
    \par This derivation shows that a Schr\"{o}dinger wavefunction, which is a solution of (4.8), is in
      fact a path amplitude. Consideration of (4.1) shows that $\psi$ is dimensionless, but in view
      of the linear nature of (4.8) it remains valid when $\psi$ is multiplied by the dimensional constant
      that is required for its physical interpretation when applied to any specific physical problem.
      \par Specialising to the spatial wavefunctions of an electron in a hydrogen atom,
       corresponding to a time-independent
      potential function, and which are interpreted according to the Born rule, the normalisation constants
      have dimension  $L^{-\frac{3}{2}}$. 
      In his original paper on transformation functions in quantum mechanics~\cite{DiracTFQM}
      Dirac made the following comment on the physical meaning of  such atomic wavefunctions (Dirac's italics):
     \par {\it The eigenfunctions of  Schr\"{o}dinger's wave equation are just the transformation
      functions (or the elements of the transformation matrix...) that enable one to transform from
      the ($q$) }[i.e. spatial coordinate]{\it scheme of matrix transformations to a scheme in which
      the Hamiltonian is a diagonal matrix.}\footnote{Wavefunctions, $\psi_i$, for which the Hamiltonian is a diagonal matrix
      are eigenfunctions of the energy: $H\psi_i = E_i\psi_i$. Then $\int \psi_i^{\ast}H\psi_jd^3x
      \equiv H_{ij} = \delta_{ij}E_i$ as a consequence of the orthonormality of the
      eigenfunctions:  $\int \psi_i^{\ast}\psi_jd^3x = \delta_{ij}$.}   
       \par This was probably the key physical insight, concerning the physical interpretation of
      Dirac's mathematical formalism, that this paper contained.
      The derivation of the Schr\"{o}dinger equation just presented shows that Dirac's assertion,
       stated in the language of Feynman's formulation of quantum mechanics, is:
      \par{\it The Schr\"{o}dinger wavefunction of an atom is the probability amplitude that an electron
      is found at a certain position in the atom when the latter is in a stationary energy state.}
        \par The probability to find the electron at a certain position  does not depend upon
      time as long as the atom remains in the same stationary state. As will be discussed in the
      following section, the situation is entirely different when path amplitudes are used to 
      describe particles moving freely in space-time. It is shown there how the appropriate path
      amplitudes for such particles may be derived directly from the transformation  equations
      (2.5) and (2.6) of Hamilton's principle function.
  \SECTION{\bf{Derivation of the path amplitude for a free particle from Hamilton's principle function}}
         The Hamilton's principle function for a free particle (denoted as $\Sc$) with energy $E$ and momentum $p$
       is obtained immediately by
      integration of the transformation equations for $S$ in (2.4 ) and (2.5):
        \begin{equation}
         \frac{\partial \Sc}{\partial x} = p,~~~ \frac{\partial \Sc}{\partial t} = -H = -E
        \end{equation}
        These equations have the evident solution:
            \begin{equation}
                 \Sc = px-Et
            \end{equation}
        where $p$ and $E$ are constants, independent of spatial position and time.
        This equation can be used as the basis, not only for the quantum mechanical description
        of a freely moving particle, but also to develop its relativistic kinematics and space-time geometry.
        If the particle, of Newtonian mass $m$, is at rest then $E = mc^2$ and the path amplitude is,
        from (4.1) and (5.2):
          \begin{equation}
         \psi(\tau) = \exp\left\{-\frac{i mc^2}{\hbar}\tau\right\}
         \end{equation}      
           where $\tau$ is the proper (rest frame) time. Since it was demonstrated, in the previous section, that
            in Feynman's formulation the conventional wavefunction {\it is} a path amplitude, the wavefunction symbol $\psi$
           is consistently employed for the path amplitude on the right side of (5.3).
           This formula, with the replacement $mc^2 \rightarrow mc^2-i\hbar/(2T)$, where $T$ is the mean lifetime of an unstable
              particle\footnote{See Section 7 below for the connection of the $T$-dependent term with the
                energy-time uncertainty relation.}, has been employed in the phenomenology
            of temporal flavour oscillations of neutral kaons since the middle 1950's~\cite{FOscNK}.
          \par The Lorentz invariant quantity $\Sc$ can be written in several equivalent ways. 
                 With a suitable choice of the origins of spatial and temporal coordinates (such that $x =0$ when
          $t=0$) the equation of motion of the particle may be written as $x = vt$. The latter may then 
         be used to eliminate $x$ from (5.2):
          \begin{equation}
              \Sc = px-Et = (pv-E)t = -E\left(1-\frac{p}{E}v\right)t.
          \end{equation}   
            In the non-relativistic limit where
           \begin{equation}
               E \simeq mc^2 +\frac{p^2}{2m},~~~p \simeq mv
              \end{equation}  
               (5.4) may be written as
             \begin{eqnarray}
    \Sc & = & -\left(mc^2+\frac{p^2}{2m}\right)\left[1-\frac{pv}{mc^2+\frac{p^2}{2m}}\right]t +{\rm O}(\beta^4)
        \\
         & = & -\left(mc^2+\frac{p^2}{2m}\right)\left[1-\frac{mv^2}{mc^2}\right]t +{\rm O}(\beta^4)
       \\
           & = & -\left(mc^2+\frac{p^2}{2m}\right)(1-\beta^2)t +{\rm O}(\beta^4)
        \end{eqnarray}
          where $\beta \equiv v/c$. With the replacements $mc^2+ p^2/2m \rightarrow E$ in (5.8) and 
            $p/E \rightarrow v/c^2$ in (5.4) the approximate equation (5.8) becomes identical to
            the exact one (5.4). Then
                   \begin{equation}
              \Sc = -E(1-\beta^2)t =-\frac{E}{\gamma}\frac{t}{\gamma} = -mc^2 \tau
          \end{equation} 
      where
      \begin{equation}
      E \equiv \gamma mc^2,~~~~~~~ p \equiv \gamma m v,~~~~~~~\tau \equiv \frac{t}{\gamma}
          \end{equation} 
          and $\gamma \equiv 1/\sqrt{1-\beta^2}$. These, together with the equation of motion
         $x = vt$, are the fundamental equations of relativistic kinematics and space-time geometry.
          It also follows from (5.9) that
                       \begin{equation}
              \Sc = -mc^2(\sqrt{1-\beta^2})t = \int -mc^2(\sqrt{1-\beta^2})dt
          \end{equation} 
                  which, making use of (2.13) identifies the non-covariant relativistic Lagrangian,
              $\Lc$, of a free particle as:
                           \begin{equation}
              \Lc \equiv -mc^2\sqrt{1-\beta^2}.
          \end{equation}        
          For consistency with quantum mechanics there can be no arbitary additive constant on the
          right side of (5.12). Planck used the Lagrangian of (5.12) (including also an arbitary
          additive constant on the right side) in his orginal derivation~\cite{PlanckRK}
         of the formulae in (5.10) for relativistic energy and momentum. This Lagrangian may also be written in
          a manifestly covariant manner as~\cite{ZZP}
          \begin{equation}
              \Lc  = -\frac{m}{2}(V^2+c^2)
          \end{equation}           
                where $V$ is the four-vector velocity: $V \equiv (\gamma c ,\gamma \vec{v})$.
           Integrating the corresponding Lagrange equation for the temporal coordinate 
           gives the time dilation relation, the last of Eqs.~(5.10)~\cite{ZZP}.
          \par The Lorentz-invariant form (5.2) for the phase of the path amplitude for a free particle
      as well as its interpretation according to Feynman's postulates were both implicit in work 
         by de Broglie published in (1924)~\cite{dBPM}. This paper also contains contains an
        interpretation of interference effects in electrodynamics in terms of probability amplitudes
       for detection of photons, as in quantum electrodynamics.
     \SECTION{\bf{The de Broglie and Planck-Einstein relations. `Wave particle duality' and
       `wave packets' for free particles}}
        The various related topics discussed in this section are either of intrinsic interest or introduce
        background material essential for the critical review of Schr\"{o}dinger's  derivations of
        quantum wave equations in Section 8 below. The topics are:
      \begin{itemize}
      \item[(i)] Derivation of the de Broglie wavelength and Planck-Einstein relations by mathematical
            subsitution in the path amplitude formula for a free particle.
     \item[(ii)] A review of  de Broglie's own derivation of the wavelength formula and his (independent)
                introduction of the concepts of the phase and group velocities of `matter waves'.
    \item[(iii)] A review of Hamilton's analogy between the classical mechanics of particle motion and
                 geometrical optics.
     \item[(iv)] A discussion of `matter waves' in the presence of electromagnetic fields leading to
                 a short path amplitude derivation of the magnetic Aharonov-Bohm effect.
   \end{itemize}
     Point (iii) is essential for the understanding of Schr\"{o}dinger's second derivation of the
     quantum wave equation. (ii) is of historical interest in its own right, while (i) and (iv) give a modern
     perspective on `wave-particle duality' within the path integral formulation. All this means inevitably
     a certain amount of repetition. For example, different derivations of the `phase velocity', $v_{\phi}$,
     are found in (i), (ii) and (iii), and of the `group velocity' $u$,
    in (i) and (ii). In fact, the same problem
    is addressed from two different historical perspectives (those of de Broglie and Hamilton) and one modern one
    ---the path integral formulation. An attempt is made in the discussion to distingush between the useful
      phenomenological concept of the de Broglie wavelength and the empty mathematical abstractions which
     are the phase and group velocities of hypothetical `matter waves'. 
         \par The path amplitude for a free particle in one dimensional motion may be written, using (5.2), as
          \begin{equation}
           \psi(x,t) \equiv \exp\left\{-\frac{i}{\hbar}\phi(x,t)\right\}
          \end{equation}
          where
           \begin{eqnarray}
           \phi(x,t) & \equiv &\frac{2 \pi}{h}(Et-px) = 2 \pi(\nu t-\frac{x}{\lambda})
        = \frac{2 \pi}{\lambda}(v_{\phi}t-x), \\
            \nu &\equiv & \frac{E}{h},  \\
            \lambda & \equiv & \frac{h}{p},  \\
             v_{\phi} & \equiv & \nu \lambda = \frac{E}{p} = \frac{c^2}{v} \ge c.
           \end{eqnarray}
       Here (6.3) is a transposition of the Planck-Einstein relation $E = h \nu$ and (6.4) defines the
       de Broglie wavelength, $\lambda$, of the particle. The energy, $E$, and the momentum, $p$, are the
       relativistic quantities defined in (5.10). The last member of (6.2) suggests the association
       of a `wave' with phase velocity $v_{\phi}$ with the path amplitude. According to the last 
       menber of (6.5) since the particle velocity $v \le c$ then $v_{\phi} \ge c$. For photons
       or other massless particles with $v = c$, then  $v_{\phi} = c$ so that the phase and particle
       velocities are the same. For massive particles the `phase velocity' defined by the purely mathematical
       substitutions in (6.3)-(6.5) is superluminal and devoid of any operational physical meaning.\footnote{
        This superluminal phase velocity of hypothetical `matter waves' should not be conflated with the real
        physical effects that have apparently been observed in several recent experiments; for example in quantum 
         tunneling~\cite{BuW,HGW} on in the near-zone of electromagnetic waves~\cite{MRR,KJAP,JHFPPNL}.}
       However, following the original
       suggestion of de Broglie, the particle velocity, $v$, is commonly associated with the `group velocity'
       of a hypothetical `wave packet' that follows from the wavelength dependence of the frequency (actually, from
       Eqs.~(6.3) and (6.4), the relativistic momentum dependence of relativistic energy) of the associated
        `phase wave'. The definitions of $E$,
       $p$ and $\gamma$ in (5.10) and (6.3)-(6.5) give:
         \begin{equation}
          m^2c^4 \equiv  m^2c^4(\gamma^2-\gamma^2\beta^2) = E^2-p^2c^2 \equiv h^2\nu_0^2 =  h^2\nu^2-\frac{h^2c^2}{\lambda^2}
         \end{equation}
            so that
        \begin{equation}
           k \equiv \frac{1}{\lambda} = \frac{\sqrt{\nu^2-\nu_0^2}}{c}.
        \end{equation}
            Differentiating (6.7) gives the `group velocity' $u \equiv d \nu/d k$:
      \begin{equation}
         \frac{1}{u} = \frac{dk}{d \nu} = \frac{1}{c} \frac{\nu}{\sqrt{\nu^2-\nu_0^2}}
         = \frac{\nu \lambda}{c^2}= \frac{v_{\phi}}{c^2}
        \end{equation}
        or 
  \begin{equation}
         u = \frac{c^2}{v_{\phi}} = v
    \end{equation}
     where (6.5) has been used. Thus the hypothetical group velocity associated with the hypothetical phase
     wave suggested by the last member of (6.2) is indeed equal to the physical velocity of the particle,
     but this does not mean, as asserted by de Broglie, and subsequently in most text books on quantum mechanics,
     that a moving particle {\it is} a wave packet and so has a dual ontological nature. In this context, there
     is an important difference, discussed in some detail in Ref.~\cite{JHFEJP}, between massive particles 
     and massless photons, for which `phase', `group' and particle velocities are all equal and for which
     `electromagnetic waves' are manifestations of beams of real photons. There is no
      equivalent, for massive particles, of the macroscopic potentials and fields $\vec{A}$, $\vec{E}$ and $\vec{B}$
      which constitute the fundamental physical concepts of classical electromagnetism.
     \par It is interesting to compare de Broglie's derivation of the formula (6.4) in his
      1927 Nobel Prize acceptance speech~\cite{deBroglieNPS} with that presented above. This calculation used
       directly the definitions of relativistic energy and momentum given in (5.10). It is first assumed
       that the phase of a particle at rest is $\phi_0 =2 \pi \nu_0 \tau$ where  $\nu_0 \equiv mc^2$ and 
       $\tau$ is the proper time of the particle. The Lorentz transformation is then used to transform the 
      proper time into a frame in which the particle is in motion with velocity $v = \beta c$ to give:
       \begin{equation} 
       \phi_0 =  2 \pi \nu_0 \tau = 2 \pi \nu_0\gamma(t-\frac{\beta x}{c}).
       \end{equation}
        The frequency transformation is given by the Planck-Einstein relation, $E = h\nu$, and the relativistic
        energy definition in (5.10), as:
       \begin{equation} 
            \nu = \gamma \nu_0.
       \end{equation}
          Inspection of (6.10) shows that the phase velocity of the associated wave
          is (compare with the last member of (6.2)):
     \begin{equation} 
          v_{\phi} = \frac{c}{\beta}
     \end{equation}
       The formula defining the `de Broglie wavelength', $\lambda$, given by $\lambda = v_{\phi}/\nu$,
     follows directly from (6.12), the definitions 
       of relativistic momentum and energy and the Planck-Einstein relation:
        \begin{equation}
         p \equiv \gamma m v =\frac{Ev}{c^2} = \frac{h \nu \beta}{c} = \frac{h \nu}{v_{\phi}} = \frac{h}{\lambda}.
        \end{equation} 
         There is, therefore, no appeal to the concept of `group velocity' in this calculation that introduces
        the measurable (and experimentally confirmed) de Broglie wavelength of a free particle. Nevertheless,
        de Broglie used (6.11), (6.12) and the `refractive index': defined as $n \equiv v/v_{\phi}$
        to derive the relation:
           \begin{equation}
           \beta = n = \sqrt{1-(\nu_0/\nu)^2} 
           \end{equation} 
            which together with a formula for the group velocity due to Lord Rayleigh:
              \begin{equation}
                  \frac{1}{u} = \frac{1}{c} \frac{d(n \nu)}{d \nu}
        \end{equation} 
                is used to show that $u = v$, so that, from (6.12), the phase and group
            velocities are related according to:
             \begin{equation} 
               u v_{\phi} = c^2.
               \end{equation} 
            \par This calculation of de Broglie is of particular interest because the formula (6.10) for the 
            $x$, $t$ dependence of the path amplitude phase, which is (up to a sign) identical to that
            of Eq.~(6.2), is here derived not from the transformation equations of S, but from a 
           relativistic time transformation that demonstrates the Lorentz invariance of the phase:
          \begin{equation} 
        \phi_0 \equiv \frac{2 \pi mc^2}{h}\tau = \frac{2\pi(Et-px)}{h} \equiv \phi
          = \frac{2\pi(E't'-p'x')}{h} \equiv \phi'.
       \end{equation} 
        This suggests, as indicated by the discussion of Eqs.~(5.6)-(5.9), that relativistic
        kinematics and space-time geometry are already somehow implicit in the formula $S = px-Et$ for Hamilton's
        principle function for a free particle. In particular, the fundamentally important relative
        minus sign between spatial and temporal components in the Minkowski metric is predicted.
        De Broglie had earlier pointed out~\cite{dBPM} this Lorentz scalar character of the path amplitude phase. 
        \par In his second 1926 paper on wave mechanics~\cite{SchrodWM2} Schr\"{o}dinger gave an extended discussion
          of the analogy between the classical mechanics of particles and geometrical optics, as suggested by Hamilton
          in 1834. In view of the connection between this discussion and  Schr\"{o}dinger's second derivation
          in the same paper of the time-independent quantum wave equation, Hamilton's ideas are
         briefly reviewed here. In the case of a time-independent Hamiltonian: $H = E$, Hamilton's principle function
        may be written as~\cite{Goldst1}:
         \begin{equation}
          S(\vec{x},t) = W(\vec{x})-Et
           \end{equation}
            where $W$, which is a function only of the spatial coordinates $\vec{x}$, is called {\it Hamilton's characterstic function}.
          A consequence of (6.18) is that surfaces of constant $S$ 
           correspond to a time evolution of $W$ according
          to $dW = E dt$. The spatial evolution of $W$, analogous to that of a wave front in physical optics, is then:
         \begin{equation}
              dW = |\vec{\nabla}W|ds
            \end{equation}
            where $ds$ is the displacement of a spatial surface of constant $W$ and 
         \begin{equation}
        \vec{\nabla} \equiv \hat{\imath}\frac{\partial~}
        {\partial x}+
        \hat{\jmath}\frac{\partial~}{\partial y}
         +   \hat{k}\frac{\partial~}{\partial z}
        \end{equation} 
              where $\hat{\imath}$,$\hat{\jmath}$,$\hat{k}$ are
              unit vectors parallel to the $x$,$y$,$z$ axes.
           A spatial surface of constant $W$  therefore moves with speed:
        \begin{equation}
          v_W = \frac{ds}{dt} = \frac{1}{|\vec{\nabla}W|}\frac{dW}{dt} =  \frac{E}{|\vec{\nabla}W|}.
          \end{equation} 
           The generalisation of the first of Eqs.~(2.5) gives
       \begin{equation}
         \vec{\nabla}S = \vec{\nabla}W = \vec{p}
          \end{equation}
       so that
            \begin{equation}
             v_W = \frac{E}{|\vec{p}|}.
          \end{equation}
            For relativistic motion of a free particle comparison of (6.23) with (6.5)
            shows that 
      \begin{equation} 
              v_W = v_{\phi} = \frac{c^2}{v}.
           \end{equation}
     For non-relativistic motion in a time-independent potential, as considered by Schr\"{o}dinger, use of
      the non relativistic ($NR$) Hamiltonian:
         \begin{equation}
       H_{NR} =  E_{NR} = \frac{p^2}{2m}+V \equiv T_{NR}+V
       \end{equation}
          gives, according to (6.23),
               \begin{equation}
             v_W(NR) = \frac{E_{NR}}{p} = \frac{E_{NR}}{\sqrt{2m(E_{NR}-V)}} =\frac{E_{NR}}{\sqrt{2m T_{NR}}}. 
          \end{equation}
        For a free particle where $E_{NR} = p^2/2m$:
                   \begin{equation}
             v_W(NR) =\frac{p^2}{2mp} = \frac{p}{2m} = \frac{v}{2}. 
          \end{equation}        
            Since the relativistic formula (6.24) is valid for all values if $v$, including those for
            which the approximation $T \simeq T_{NR} = p^2/(2m)$ is a good one, the inconsistency between
              the value of $v_W$ and $v_W(NR)$ is an indication of the lack of any operational
            physical meaning for the hypothetical `phase velocities' $v_{\phi}$, $v_W$ and  $v_W(NR)$ within 
            classical mechanics. The relativistic Hamiltonian $H$ differs from  $H_{NR}$  by the 
             inclusion of the rest-mass contribution $mc^2$. As is well-known, arbitary additive constants
            in Hamiltonians or Lagrangians leave unchanged all physical predictions in
             classical mechanics. As will be discussed Section 8 below, Schr\"{o}dinger used (6.26) 
             together with a classical wave equation in his second published derivation of
             the time-independent quantum-mechanical wave equation.
             \par Within quantum mechanics, the formula (6.27) can be derived directly from the
               formula (6.2) for $\phi(x,t)$ without any consideration of Hamilton's analogy between
               classical mechanics and geometrical optics: 
              \begin{eqnarray}
             \phi & = & -\frac{2 \pi}{h}(Et-px) = -\frac{2 \pi}{h}(\gamma mc^2 t-\gamma mxv) \nonumber \\
                  &  = & -\frac{2 \pi}{h}\left[\left( mc^2+ \frac{mv^2}{2}\right)t-mvx\right]+{\rm O}(\beta^2) \nonumber \\
                  &  = & \phi_m +\phi_{NR}+{\rm O}(\beta^2)  
           \end{eqnarray}
               where
            \begin{equation}
              \phi_m \equiv -\frac{2 \pi mc^2}{h}t,~~~\phi_{NR} \equiv  -\frac{2 \pi p}{h}\left(\frac{v}{2}t-x\right)
                             = -\frac{2 \pi}{\lambda}(v_W(NR)t-x).
             \end{equation}
              \par The unphysical nature of the  `phase velocities' $v_{\phi}$, $v_W$ and  $v_W(NR)$ is to be
               contrasted with that of the measurable de Broglie wavelength of a free particle
               for which, unlike, $v_W$ and  $v_W(NR)$, $\lambda$ and  $\lambda_{NR}$ have consistent values:
       \begin{equation}
          \lambda = \frac{h}{p} =  \frac{h}{\gamma mv} =\frac{h}{mv}  +{\rm O}(\beta^2) 
        \equiv  \lambda_{NR}  +{\rm O}(\beta^2)
       \end{equation}
           while
     \begin{equation}
    \frac{v_W(NR)}{v_W} = \frac{v/2}{c^2/v} = \frac{1}{2} \beta^2 .
     \end{equation}
      So that the {\it ratio} of $v_W$ and  $v_W(NR)$, not their difference, is of O($\beta^2$)!
     \par The unphysical nature of the `phase velocity' of a massive particle and the associated `phase wave' 
        is further illustrated 
          by considering the generalisation of de Broglie's `refractive index' in Eq.~(6.14) to the case of the
         motion of a charged particle in electric and magnetic fields~\cite{ES}. It is found that not only 
         the `phase velocity' but also the `wavefront' directions as well as the associated `wavelength'
          all depend on the gauge function $\xi$ used the specify the vector potential according to
          $\vec{A}(\xi) \equiv \vec{A}_0+\vec{\nabla}\xi$. The phase of the path amplitude of a particle of
          electric charge $q$, moving from $A$ to $B$ along a path $j$ in a region of non-vanishing vector
         potential is
         \begin{eqnarray}
          \phi^{(j)}_{AB} & = & \int_A^B[ L^{(j)}_{{\rm kin}}+ q \vec{v} \cdot \vec{A}^{(j)}(\xi)] dt
           \nonumber \\
       & \equiv &  - S^{(j)}_{AB,{\rm kin}} + q\int_A^B \hat{s} \cdot \vec{A}^{(j)}(\xi) ds
           \nonumber \\
       & = &  - S^{(j)}_{AB,{\rm kin}} + q\int_A^B \hat{s} \cdot \vec{A}^{(j)}_0 ds+
          q\int_A^B \hat{s} \cdot \vec{\nabla}\xi^{(j)}ds 
           \nonumber \\
  & = &  - S^{(j)}_{AB,{\rm kin}} + q\int_A^B \hat{s} \cdot \vec{A}^{(j)}_0 ds+
          q(\xi_B-\xi_A) 
          \end{eqnarray}
           where $\hat{s}$ is a unit vector along the path of the particle, and $L^{(j)}_{{\rm kin}}$ is the 
          free-particle Lagrangian of Eq.~(5.12). Thus the physically measureable
           {\it phase difference} between two paths with fixed start and end points, is, unlike the phase velocity,
           wavefront direction or wavelength, independent of $\xi$ and therefore gauge invariant. Considering two
          paths with equal values of $ S_{AB,{\rm kin}}$, (6.32) gives:  
               \begin{eqnarray}
              \Delta \phi_{AB} & = & \phi^{(l)}_{AB}- \phi^{(j)}_{AB} = q\left[\int_A^B  \hat{s}\cdot \vec{A}^{(l)} ds
                 - \int_A^B  \hat{s}\cdot \vec{A}^{(j)} ds \right] \nonumber \\
               & = & q\int_S\hat{n} \cdot \vec{H}dS = q f_H  
             \end{eqnarray}
                where in the last member Stoke's theorem has been used and $f_H$ is total flux of the magnetic
                field $\vec{H}$ threading the area between the paths $l$ and $j$. The phase shift $\Delta \phi_{AB}$
                in (6.33) is the magnetic Aharonov-Bohm effect~\cite{AB}, as actually first pointed out by
                Ehrenberg and Siday~\cite{ES}
               \par In conclusion, the `waves' commonly associated with the path amplitudes, defined over macroscopic
                 distances, of massive particles moving either freely or in fields of force,
               are a mathematical abstraction that
                 unlike the electromagnetic waves associated with massless photons, or the path amplitudes themselves,
                  are devoid of any
                 operational physical significance. Also meaningless is the ontological concept of `wave particle
                duality'. What exist in the physical world are the particles. Following de Broglie, the probability
                amplitudes of quantum mechanics, which are purely mathematical in nature, have been historically
                conflated  with conjectured `matter waves'. Indeed, as shown above, there is an exact correspondence
                between the space time functionality of quantum path amplitudes and that of the physically-existing,
                but completely independent,
                waves of classical physics. This is a mathematical accident devoid of any physical significance.
              To summarise: only the particles exist physically, the purely mathematical probability
                amplitudes\footnote{They are mathematical constructs analogous to potentials in
                  classical mechanics which clearly have an ontological status quite different to that
                  of the physical objects, the motion of which, they encode.} tell how they propagate in space-time.
                  The only `wave like' parameter of phenomenological
                  significance is the de Broglie wavelength of a free particle, which, unambigously determined
                 by the values of its momentum and Planck's constant, is a useful, but not fundamental,
                 physical quantity. It was, of course, natural,
                historically, and before the advent of quantum
                mechanics, to associate `waves' with light, in view of the simple and quantitatively correct
                 explanations of the interference experiments of Young and Fresnel provided by the photon de Broglie
                 wavelength concept, within the classical wave theory of light. 
                  \par It is not generally known that the first quantitative quantum
                 mechanical experiment was performed, not in the 19th or early 20th Century, but some
                 two hundred years earlier, by Newton, in his analysis of the structure of interference
                patterns produced in the thin air film between a flat glass plate and a convex lens
              ---Newton's Rings~\cite{NR}. It is demonstrated in Ref.~\cite{JHFAP} that
                the classical wave theory of light, in which only the spatial component of the wave
               is considered, is a necessary consequence of the Feynman path integral description~\cite{FeynQED,JHFAP} of 
               the production of a photon by an excited atom and its subsequent detection.

        \SECTION{\bf{Eigenvalues, eigenstates, canonical commutation relations and the 
          time-independent Schr\"{o}dinger equation}} 
         Considering the wavefunction or path amplitude for a free particle in one-dimensional
         motion, it follows from (6.1) and (6.2) that:
            \begin{equation}
           -i\hbar \frac{\partial \psi(x,t)}{\partial x} = p \psi(x,t) 
            \end{equation}
             which may be generalised to three spatial dimensions by making the making the replacement:
    \[ px \rightarrow \vec{p} \cdot \vec{x} = p_xx+p_yy+p_zz \]
              in (6.2) to obtain:
           \begin{equation}
              -i\hbar \frac{\partial \psi}{\partial x} = p_x \psi,~~~-i\hbar \frac{\partial \psi}{\partial y} = p_y \psi,
       ~~~-i\hbar \frac{\partial \psi}{\partial z} = p_z \psi.
            \end{equation}
            The function $\psi(\vec{x},t)$ is therefore an {\it eigenfunction} of the differential operator
             $-i\vec{\nabla}$ with {\it eigenvalue} $\vec{p}$. Similarly since
              \begin{equation}
           i\hbar \frac{\partial \psi(\vec{x},t)}{\partial t} = E \psi(\vec{x},t)
            \end{equation}
              $\psi(\vec{x},t)$ is an eigenfunction of the differential operator $i\hbar \partial/\partial t$
               with eigenvalue $E$. Introducing symbols for the differential operators:
             \begin{equation}
              {\cal \vec{P}} \equiv -i\hbar\vec{\nabla},~~~{\cal E} \equiv i\hbar \frac{\partial ~}{\partial t}
                \end{equation}
                the {\it eigenvalue equations} (7.2) and (7.3) are:
               \begin{equation} 
     \vec{{\cal P}}\psi(\vec{x},t;\vec{p},E) = \vec{p}\psi(\vec{x},t;\vec{p},E),~~~
          {\cal E}\psi(\vec{x},t;\vec{p},E) = E\psi(\vec{x},t;\vec{p},E)
            \end{equation}
         where the eigenfunction $\psi$ is labelled by its eigenvalues. These equations may be contrasted with
      the case of a particle moving under the influence of a non-vanishing potential $V$, for example
    the electron in a hydrogen atom. In this case, the electron wavefunctions are eigenstates of the
    energy of the {\it atom} while the electron itself is neither in an eigenstate of momentum nor of energy.
       See the detailed discussion around Eqs.~(7.18)-(7.28) below.
         \par It is important to notice here
          that, in order that  $\psi$ be an eigenfunction it is {\it necessary that its space-time dependence
         is a complex exponential} with  phase as in (6.2). Replacing the complex exponential function
          by a real harmonic function such as $\cos[(\vec{p} \cdot \vec{x}-Et)/\hbar]$ will not give the
         eigenvalue equations (7.5). Here the appearance, in the equations of quantum mechanics, of
         $\sqrt{-1}$ is mandatory, if the concepts of eigenfunctions and eigenvalues are to introduced.   
          \par In view of the definitions of $\vec{{\cal P}}$ and ${\cal E}$,  the following identities
           hold, where $f$ and $F$ are arbitrary functions of $x$ and $t$, as a consequence of the
           product rule of differential calculus:
            \begin{eqnarray}
             \left({\cal P}_xf-f{\cal P}_x+i\hbar \frac{\partial f}{\partial x}\right)F & \equiv & 0, \\
             \left({\cal E}f-f{\cal E}-i\hbar \frac{\partial f}{\partial t}\right)F & \equiv & 0.
            \end{eqnarray}
             To verify (7.6) the definition: ${\cal P}_x \equiv -i\hbar\partial/\partial x$ is substituted on the
             left side to give:
                  \begin{eqnarray}
            {\cal P}_xfF-f{\cal P}_xF+i\hbar \frac{\partial f}{\partial x}F & = & 
                    -i\hbar\left( \frac{\partial (fF)}{\partial x} -f\frac{\partial F}{\partial x} -
                    \frac{\partial f}{\partial x}F\right) \nonumber \\
                 & = &   -i\hbar\left( \frac{\partial f}{\partial x}F+f\frac{\partial F}{\partial x}
                    -f\frac{\partial F}{\partial x} - \frac{\partial f}{\partial x}F\right) \nonumber \\
                  & = & 0.  \nonumber            
            \end{eqnarray}
             Setting $f = x$ or $f = y$, (7.6) is conventionally written, in a symbolic manner, on
            cancelling the arbitrary factor $F$ on both sides of the equation:
                \begin{eqnarray} 
             x{\cal P}_x-{\cal P}_xx & = & [x,{\cal P}_x] = i\hbar \\
            y{\cal P}_x-{\cal P}_xy & = & [y,{\cal P}_x] = 0
           \end{eqnarray}
            where the commutator $[A,B] \equiv AB-BA$ is introduced. 
             With the notation:
            \[ (x,y,z)\equiv (x_1,x_2,x_3),~~~~({\cal P}_x,{\cal P}_y,{\cal P}_z) \equiv 
        ({\cal P}_1,{\cal P}_2 ,{\cal P}_3)  \]
          (7.8) and (7.9) generalise to
                \begin{equation} 
             x_j {\cal P}_{k}-{\cal P}_{k}x_j  =  [x_j,{\cal P}_{k}] = i\hbar\delta_{jk}~~~j,k =1,2,3
           \end{equation}
             where $\delta_{jk}$, the Kronecker $\delta$-function is unity when $j = k$ and zero otherwise.
             The relations (7.10) called {\it canonical commutation relations} or {\it 
              fundamental quantum conditions}~\cite{DiracFQC} played a key role in the conceptual 
              development of the matrix mechanics version of quantum mechanics. However, it should not
              be forgotten that due to the presence of the term  $x_j {\cal P}_{k}$, which is
              a differential operator, the relations are mathematically meaningless 
              unless multiplied on the right by some function of $x_1$, $x_2$ and $x_3$. Also the meaning of
               the term
              ${\cal P}_{k}x_j$ depends also on the presence of the function $F$, because the
             differental operator ${\cal P}_{k}$ in general acts not only 
              on $x_j$ i.e. to  give ${\cal P}_{k}(x_j)$, corresponding to $F = 1$, but implictly
              on the omitted arbitary function $F$ to give instead ${\cal P}_{k}(x_j F)$. In all actual
             physical applications of (7.10), for example, derivations of the Heisenberg uncertainty
             relation~\cite{Robertson}, or the eigenvalues of the harmonic oscillator~\cite{DiracHO},
              or calculation of atomic transition matrix elements~\cite{SakuraiATM} the relation
              (7.10) is indeed multiplied on the right by spatial wavefunctions that do depend on 
               $x_1$, $x_2$ and $x_3$.
            \par Dirac~\cite{DiracFQC,DiracPRSA} introduced quantum commutation relations for arbitary 
              canonically-conjugate variables, $u$, $v$ by postulating {\it a priori} the following connection
         with a corresponding Poisson bracket, $\{u,v\}$, of classical mechanics:\footnote{This is the place
         (on p84!) in Dirac's book on quantum mechanics~\cite{DiracFQC} where Planck's constant makes its
         first appearence.}
            \begin{equation}
             [u,v] \equiv i\hbar\{u,v\} 
            \end{equation}
             where, for a system with spatial coordinates $q_r$ and momenta $p_r$~\cite{GoldPB}:
        \begin{equation}
            \{u,v\} \equiv \sum_r\left(\frac{\partial u}{\partial q_r}\frac{\partial v}{\partial p_r}
                     - \frac{\partial u}{\partial p_r}\frac{\partial v}{\partial q_r}\right)
           \end{equation}
           The equation (7.11) is a fair copy of Dirac's one in~\cite{DiracFQC}, except that the latter wrote explicitly
               the commutator as $uv-vu$ and used the notation  $[u,v]$ for the Poisson bracket. However it is clear from
         reading the accompanying text that the symbols $u$ and $v$ have different meanings on the left and right sides
        of the equation. On the left side they represent as-yet-unspecified quantum-mechanical operators, while on the
       right side they are the corresponding classical-mechanical quantities. Only later is the
        classical $\leftrightarrow$ quantum  correspondence:  $\vec{p} \leftrightarrow {\cal \vec{P}} \equiv -i\hbar\vec{\nabla}$
        suggested.
           If $u$ and $v$ are a Cartesian cordinate, $x_j$, and momentum component, $p_k$, respectively and $q_r =  x_r$, then (7.12) gives:
        \begin{equation}
            \{x_j,p_k\} = \sum_r\left(\frac{\partial x_j}{\partial x_r}\frac{\partial p_k}{\partial p_r}
                     - \frac{\partial x_j}{\partial p_r}\frac{\partial p_k}{\partial x_r}\right) 
                = \sum_r\left(\frac{\partial x_j}{\partial x_r}\delta_{kr}\right) = \sum_r \delta_{jr}\delta_{kr} =
                     \delta_{jk}               
           \end{equation}
           The Poisson bracket on the right side of (7.11), in the interesting case when $u \equiv  x_j$
           and $v \equiv  p_j$ and $q_r = x_r$, is then nothing more than a very complicated way to write the Kronecker
          $\delta$-function $\delta_{jk}$ on the right side of (7.10) above! This means that Dirac's postulate (7.11)
          is equivalent to the relation (7.10) in this case. However (7.10) is not postulated in the present paper
          but {\it derived} from the
            path amplitude of a free particle, (6.1) and (6.2). 
           \par By inverting the mathematical manipulations leading from (7.1) to (7.10) above it is possible
              to start with (7.11) as initial premise and arrive at (7.1). Just this approach is the one followed
             in Chapter IV of Dirac's book~\cite{DiracFQC}. The operator relation
             ${\cal\vec{P}} \equiv -i\hbar\vec{\nabla}$ of Eq.~(7.4) is derived as a consequence of (7.11). Similarly,
             in the following Chapter V of~\cite{DiracFQC} the constant with dimensions of action, $\hbar$, is introduced
             for a second time in the equation:
  \begin{equation}
          i \hbar\frac{\partial \psi}{\partial t} = H \psi 
   \end{equation}
           where $H$ is the Hamiltonian.
      As shown below, on making the substitution $\vec{p} \rightarrow \vec{P} \equiv -i\hbar\vec{\nabla}$
                 for the momentum argument of $H$, (7.14) becomes the time-dependent Schr\"{o}dinger equation. This is how
       the latter is introduced in Ref.~\cite{DiracFQC}. Feynman's postulate II: $\psi = \exp[iS/\hbar]$ finally appears on
        p121 of Ref.~\cite{DiracFQC}.\footnote{Dirac actually replaces
              $\psi$ with $\psi/A$ where A is an arbitary real function of the space-time coordinates, so that Dirac's
            $\psi$ is actually the propagator $K$ of Eq.~(3.3) above.} in a section with the title `The motion of wave packets'.
        The presentation of the subject matter in  Ref.~\cite{DiracFQC} is an accurate representation of Dirac's own
          road to discovery but, from a modern perspective, 
          has serious pedagogical shortcomings. Feynman's postulate II (actually first proposed
               for free particles by de Broglie~\cite{dBPM}, and later, in complete generality,
              by Dirac himself~\cite{DiracLQM})
                  with its clearly specified physical meaning, immediate introduction of the
                      fundamental constant $\hbar$ and its direct connection, via Hamilton's principle, with
             classical mechanics~\cite{DiracLQM,FeynRMP,EFT1,EFT2} is clearly an infinitely preferable initial postulate
           to Dirac's somewhat arcane premise (7.11).
           \par Similarly, setting $f = t$  in (7.7) the energy-time commutation relation:
         \begin{equation} 
             {\cal E}t-t{\cal E} = i\hbar
           \end{equation}
              is obtained by symbolically cancelling the arbitrary factor $F$ on both sides. By the method of
           Ref.~\cite{Robertson} this relation may be used to derive an energy-time uncertainty
            relation\footnote{Although the mathematics of the derivation of this equation by the method of
                 Ref.~\cite{Robertson} is similar to that of the more familiar space-momentum uncertainty relation:
   $\Delta x \Delta p \ge \hbar$, its physical meaning is more controversial. See~\cite{RWF} and references therein.}:
     \begin{equation} 
      \Delta t \Delta E \ge \hbar.
     \end{equation}
       An application of this is to the phase of the path amplitude of an unstable
       particle, as discussed after Eq.~(5.3) above, where $ \Delta t$ is identified with the 
       mean lifetime, $T$, of the particle and  $\Delta E$ with the Breit-Wigner decay width $\Gamma$, so that 
        the equality in (7.16) gives:   
   \begin{equation} 
        mc^2 \rightarrow mc^2-i\frac{\hbar}{2T} = mc^2-i\frac{\Gamma}{2}.
       \end{equation}
             \par A particle moving in free space is in an eigenstate of both energy and momentum.
              This may be contrasted with the case for a bound particle, for example the electron or proton
              in a hydrogen atom.  The {\it atom} is in an eigenstate of
              both energy and momentum (for example at rest) but the electron has a variable momentum
              and position. These distributions are connected by a Fourier transform that predicts
              that their widths are related by a Heisenberg uncertainty relation. The problem of
              determining the spatial wavefunctions (eigenfunctions
              describing the position of the electron relative to the
              proton, but with fixed atomic, not electron  or proton,
              energies) first solved by Schr\"{o}dinger,
               will now be considered from the
              viewpoint of Feynman's path integral formulation of
              quantum mechanics. This analysis is important for the
              comparison with Schr\"{o}dinger's second derivation of
              the quantum wave equation that is critically reviewed
              in the following section. 
              \par Since the potential function, and hence the Hamiltonian, for the hydrogen atom in
              a bound state $i$ is
              time-independent, Hamilton's principle function is given by (6.18):
                \begin{equation}
                S_{{\rm H},i} =  W_{{\rm H},i}(\vec{x})-E_{H,i}t
              \end{equation}
               where $W_{{\rm H},i}(\vec{x})$ is the corresponding  Hamilton's
               characteristic function. The wavefunction of the hydrogen atom in the $i$th bound state is then:
                \begin{equation}
 \psi_{{\rm e}}(\vec{x},t; E_{{\rm H},i}) = \exp\left[\frac{i S_{{\rm H},i}}{\hbar}\right] =
  \exp\left[\frac{i}{\hbar}(W_{{\rm H},i}(\vec{x})-E_{H,i}t)\right]
 \end{equation}
               Writing explicitly the different contributions
              to  $E_{H,i}$ gives:
             \begin{equation}
              H_{{\rm H},i} = E_{{\rm H},i} = E_{{\rm e},i}+ E_{{\rm p},i} +V =
             m_{{\rm e}}c^2 +T_{{\rm e},i}+  m_{{\rm p}}c^2 +T_{{\rm p},i} +V
             \end{equation}
               where $m_{{\rm e}}$, $m_{{\rm p}}$ are the Newtonian electron and proton masses and
              $T_{{\rm e},i}$,  $T_{{\rm p},i}$ their kinetic
                 energies in the $i$th bound state. Introducing the (negative) bound state energy,
                  or binding energy, $E_{{\rm H},i}^{{\rm BS}}$,
                 defined as:
                  \begin{equation}
      E_{{\rm H},i}^{{\rm BS}} \equiv  E_{{\rm H},i}-M_0c^2 = T_{{\rm e},i}+T_{{\rm p},i}+V,
           ~~~~M_0c^2 \equiv m_{{\rm e}}c^2+ m_{{\rm p}}c^2
                  \end{equation}
                  (7.19) can be written as:
           \begin{eqnarray}
             \psi_{{\rm e}}(\vec{x},t; E_{{\rm H},i}) & = & \exp\left[\frac{i S_{{\rm H},i}}{\hbar}\right]
                   \nonumber \\
             & = & \exp\left[\frac{-i M_0 c^2 t}{\hbar}\right]\exp\left\{\frac{i}{\hbar}
                   (W_{{\rm H},i}(\vec{x})-E_{{\rm H},i}^{{\rm BS}}t)\right\}
                \nonumber \\
                & \equiv & \exp\left[\frac{-i M_0 c^2 t}{\hbar}\right]\psi_{{\rm e}}
                (\vec{x},t; E_{{\rm H},i}^{{\rm BS}}). 
             \end{eqnarray}
         Differentiating $\psi_{{\rm e}}(\vec{x},t; E_{{\rm H},i}^{{\rm BS}})$ twice w.r.t. $\vec{x}$ , using the
          first transformation equation in (2.5), gives
         \begin{equation}
          \vec{\nabla}^2 \psi_{{\rm e}}(\vec{x},t; E_{{\rm H},i}^{{\rm BS}}) =-\frac{p_{{\rm e}}^2}{\hbar^2}
         \psi_{{\rm e}}(\vec{x},t; E_{{\rm H},i}^{{\rm BS}})
         \end{equation}
          Since the spatial wavefunction $\psi_{{\rm e}}(\vec{x}; E_{{\rm H},i}^{{\rm BS}})$ is given by
           (7.22) as
          \begin{equation} 
           \psi_{{\rm e}}(\vec{x}; E_{{\rm H},i}^{{\rm BS}}) \equiv \exp\left[\frac{i}{\hbar}W_{{\rm H},i}(\vec{x})\right]
           =\psi_{{\rm e}}(\vec{x},t; E_{{\rm H},i}^{{\rm BS}})
           \exp\left[i\frac{E_{{\rm H},i}^{{\rm BS}}t}{\hbar}\right]
          \end{equation}
            a factor $\exp[-i E_{{\rm H},i}^{{\rm BS}}t/\hbar]$ may be cancelled from both sides
            of (7.23) giving, on transposing the resulting equation:
            \begin{equation}
           p_{{\rm e}}^2 = - \hbar^2 \frac{ \vec{\nabla}^2 \psi_{{\rm e}}(\vec{x}; E_{{\rm H},i}^{{\rm BS}})}
                         {\psi_{{\rm e}}(\vec{x}; E_{{\rm H},i}^{{\rm BS}})}.
            \end{equation}
                    In the non-relativistic (NR) approximation, appropriate to the motion of the electron and proton
         in a hydrogen atom bound state, $T \simeq p^2/(2m)$ and, since the momenta and the electron and
        proton are equal and opposite in the rest frame of the atom:
           \[ \frac{T_{{\rm p},i}}{T_{{\rm e},i}} = \frac{m_{{\rm e}}}{m_{{\rm p}}} \ll 1. \]
         The proton kinetic energy term in  (7.21) may therefore be neglected, in first appreoximation, so that
         \begin{equation}
         H_{{\rm H},i}^{{\rm BS,NR}} = E_{{\rm H},i}^{{\rm BS,NR}} = \frac{p_{{\rm e}}^2}{2m_{{\rm e}}}+V.
          \end{equation}
            Setting $E_{{\rm H},i}^{{\rm BS}}$ equal to $E_{{\rm H},i}^{{\rm BS,NR}}$ in (7.25) and subsituting
               for $p_{{\rm e}}^2$ in (7.26) using the resulting equation, gives, on rearrangement, the {\it time-independent Schr\"{o}dinger equation}
             for the spatial wavefunction of an electron in a bound state of the hydrogen atom:
            \begin{equation}
    \vec{\nabla}^2 \psi_{{\rm e}}(\vec{x}; E_{{\rm H},i}^{{\rm BS,NR}}) +\frac{2 m_{{\rm e}}}{\hbar^2}
              ( E_{{\rm H},i}^{{\rm BS,NR}}-V)\psi_{{\rm e}}(\vec{x}; E_{{\rm H},i}^{{\rm BS,NR}}) = 0.
             \end{equation} 
              Transposing (7.24) and differentiating w.r.t. $t$ gives
         \begin{equation}
          \frac{\partial  \psi_{{\rm e}}(\vec{x},t; E_{{\rm H},i}^{{\rm BS}})}{\partial t}
        = -\frac{i E_{{\rm H},i}^{{\rm BS}}}{\hbar}\psi_{{\rm e}}(\vec{x},t; E_{{\rm H},i}^{{\rm BS}}).
          \end{equation}
          Substituting $\psi_{{\rm e}}(\vec{x},t; E_{{\rm H},i}^{{\rm BS}})$, and its time derivative from
          (7.28), 
          in the time dependent Schr\"{o}dinger equation (4.8),  cancelling out a common
           multiplicative factor $\exp[-iE_{{\rm H},i}^{{\rm BS}}t/\hbar]$, and setting
            $E_{{\rm H},i}^{{\rm BS}}$ equal to $E_{{\rm H},i}^{{\rm BS,NR}}$, the time-independent equation
          (7.27) is recovered. Notice that the NR approximation is implicit in the H-J equation
          (4.7) from which (4.8) is derived.

            \par Eq.(7.27), containing the spatial electron wavefunction and the (negative) bound state
           energy $E_{{\rm H},i}^{{\rm BS}}$, is just the partial differential equation from which
           Schr\"{o}dinger derived the bound state energies and wavefunctions of the hydrogen
           atom~\cite{SchrodWM1}. Schr\"{o}dinger's own derivations of this equation and of the
           time-dependent equation (4.8) are discussed in the following section.
          \SECTION{\bf{Schr\"{o}dinger's derivations of quantum wave equations}}
           Schr\"{o}dinger's first published derivation of the time independent-quantum wave equation
          (7.27)~\cite{SchrodWM1} considers the H-J equation of a system with a time-independent
            Hamiltonian so that
            \begin{equation} 
             \frac{\partial  S}{\partial t} = -H = -E
    \end{equation}
              where $E$ is constant, so that the H-J equation (2.6) is written:
              \begin{equation} 
               H(q_i, \frac{\partial  S}{\partial q_i}) = E 
     \end{equation}
          Schr\"{o}dinger then introduces the following ansatz relating the `wavefunction' $\psi$ (which is here introduced 
          for the first time) to Hamilton's characteristic function, denoted here as $S$, rather than $W$, as
          is conventional, and done in Sections 6 and 7 above:
         \begin{equation} 
                S = K\ln \psi
       \end{equation}
              or
 \begin{equation} 
          \psi = \exp\left[\frac{S}{K}\right]
    \end{equation}
              where K is a real constant, which, like $h$, has the dimensions of action.
           This ansatz is the same as the fundamental postulate II of Feynman's formulation of quantum
           mechanics, for the spatially-dependent part of the path amplitude, on making the replacement
           $K \rightarrow -i\hbar$. 
           \par Using (8.2) and (8.3) and following the same chain of arguments from which the
             time-dependent Schr\"{o}dinger equation (4.8) is derived from the H-J equation (4.7) 
              and Feynman's postulate (4.2) instead of (8.3), yields the equation:
                       \begin{equation}
            \vec{\nabla}^2 \psi -\frac{2 m_{{\rm e}}}{K^2}
              ( E-V)\psi = 0
             \end{equation}
        which resembles the time-independent Schr\"{o}dinger equation (7.27) except that $\hbar$ is replaced
           by $K$ and the second term on the left side has a minus sign instead of a plus sign. Now
          (8.5) is a necessary mathematical consequence of (8.2) and (8.3) when $H$ has the form of the
          non-relativistic Hamiltonian of Eq.~(4.7), but (8.5) is {\it not} the time-independent
          Schr\"{o}dinger equation, and would not, if solved, give the correct bound state wavefunctions
          and energies of the hydrogen atom. In order to derive the correct equation (7.27) starting from
           (8.2), (8.3) and the Hamiltonian of Eq.~(2.7) Schr\"{o}dinger had to introduce a further ansatz
            concerning the wavefunction $\psi$. Differentiating (8.3) to obtain equations analogous to
           (4.3) and substituting for $\partial  S/\partial x_i$ in (8.2) gives the equation
                      \begin{equation}
   \left( \frac{\partial  \psi}{\partial x}\right)^2 + \left( \frac{\partial  \psi}{\partial y}\right)^2
 + \left( \frac{\partial  \psi}{\partial z}\right)^2 -\frac{2 m_{{\rm e}}}{K^2}( E-V)\psi^2 = 0
   \end{equation}
    The quantity:
      \begin{equation}
    J \equiv \int\int\int \left[\left( \frac{\partial  \psi}{\partial x}\right)^2 + 
      \left( \frac{\partial  \psi}{\partial y}\right)^2 + \left( \frac{\partial  \psi}{\partial z}\right)^2
     -\frac{2 m_{{\rm e}}}{K^2}( E-V)\psi^2\right]dx dy dz
       \end{equation}
 is now introduced and the condition is imposed that {\it $J$ should be stationary for arbitary variations of the
      wave function $\psi$ : $\delta J = 0$.} 
      Now
       \begin{equation}
       \delta \left( \frac{\partial  \psi}{\partial x}\right)^2 = 2 \frac{\partial  \psi}{\partial x}
     \frac{\partial (\delta \psi)}{\partial x}
     \end{equation} 
      Integrating by parts:
       \begin{equation}
      \int_{x_L}^{x_U} \delta \left(\frac{\partial  \psi}{\partial x}\right)^2 dx
        = 2\left[\frac{\partial  \psi}{\partial x}\delta \psi\right]_{x_L}^{x_U}-
              2 \int_{x_L}^{x_U}\frac{\partial^2  \psi}{\partial x^2}\delta \psi dx 
     \end{equation}
         also
       \begin{equation}
     \delta (\psi)^2 = 2 \psi \delta \psi
      \end{equation}
        Substituting (8.9) and similar formulae for the $y$ and $z$ coordinates, and (8.10), into the expression for
         $\delta J$ and assuming that $\delta \psi$ vanishes at the limits of integration in the first term
            on the right side of (8.9) and the similar formulae for the other spatial coordinates, gives:
          \begin{equation}
       \frac{\delta J}{2} = 
        -\int\int\int dx dy dz \left[\vec{\nabla}^2 \psi +\frac{2 m_{{\rm e}}}{K^2}(E-V)\psi \right]\delta \psi = 0 
      \end{equation}
      Since this equation must hold for arbitary $\delta \psi$ then
       \begin{equation}
        \vec{\nabla}^2 \psi +\frac{2 m_{{\rm e}}}{K^2}(E-V)\psi = 0
         \end{equation}
    which is just the time-independent Schr\"{o}dinger equation (7.27) on setting $K = \hbar$.
     \par Schr\"{o}dinger's notes~\cite{MooreSN} show that he was well aware that the solution of (8.12) gives
       correctly the bound state energies of the hydrogen atom before introducing in~\cite{SchrodWM1} the anstaz concerning
        the hypothetical quantity $J$. This artifice compensates for the physically incorrect ansatz (8.3). The 
         constant $K$ should actually be the pure imaginary quantity $-i\hbar$ in which case (8.5) beomes the
        correct equation (7.27). Repeating Schr\"{o}dinger's stationarity algorithm starting with the correct
         relation $S = -i\hbar\ln \psi$ would then give the incorrect equation (8.5)!  Indeed,                                                                                          since the H-J equation
         and the properties of the generating function $S$ already follow from Hamilton's equations which in turn
         are a consequence of Hamilton's Principle --- the condition that the action S should be stationary
        for arbitary variations of space-time paths--- it would seem that Schr\"{o}dinger is attempting here
          to close a door that is already shut.
           \par In fact   the time dependent Schr\"{o}dinger equation (4.7), and hence (8.12), can be derived from
          a stationarity requirement imposed on an action function
            derived from a certain Lagrangian under arbitary variations of the wavefunction $\psi$. The appropriate
           Lagrangian, given by Heisenberg, is~\cite{HeisLagr}:
               \begin{equation}
              L = -\frac{\hbar^2}{2 m_{{\rm e}}}\nabla \psi^{\ast} \cdot \nabla \psi-\frac{\hbar}{2 i}
               \left(\frac{\partial  \psi}{\partial t}\psi^{\ast}- \frac{\partial  \psi^{\ast}}{\partial t}\psi\right)
                   +eV \psi\psi^{\ast}+\frac{1}{8\pi}\nabla V \cdot \nabla V. 
              \end{equation}
               Requiring that the corresponding action is stationary for variations of $\psi$ or $\psi^{\ast}$ derives
               the time-dependent Schr\"{o}dinger equation (4.7), or its complex conjugate, respectively. Similarly
            varying $V$ gives Poisson's equation:
               \begin{equation}
                \nabla^2 V = -4 \pi(\rho+\rho_0)
        \end{equation}
       where $\rho \equiv -e\psi\psi^{\ast}$ and $\rho_0$ represents electric charges other than that of the
        electron described by $\psi$. 
        \par As already mentioned, Schr\"{o}dinger's second paper on wave mechanics~\cite{SchrodWM2} discusses
        Hamilton's analogy between classical mechanics and geometrical optics, via the H-J equation that
        Schr\"{o}dinger calls the `Huygens' Principle'. Essentially the same argument as that given above in Section 6
         leads to the formula (6.26) for the phase velocity of the waves. In~\cite{SchrodWM2} the time-independent
        equation (8.12) is then rederived from the corresponding classical wave equation:
             \begin{equation}
         \nabla^2 \psi -\frac{1}{v_W(NR)^2}\frac{\partial^2  \psi}{\partial t^2} = 0.
              \end{equation}
          On the assumption the $\psi$ has harmonic time dependence with frequency $\nu$ then:
                   \begin{equation}
        \frac{\partial^2  \psi}{\partial t^2} = -4 \pi^2 \nu^2 \psi. 
              \end{equation}  
            Assuming also the Planck-Einstein relation (6.3) then combining (6.26), (8.15) and (8.16) gives:
          \begin{equation}     
        \nabla^2 \psi -\frac{1}{v_W(NR)^2}\frac{\partial^2  \psi}{\partial t^2} =
          \nabla^2 \psi -\frac{2 m_{{\rm e}}(E-V)}{h^2 \nu^2}\frac{\partial^2  \psi}{\partial t^2}
              =  \nabla^2 \psi+\frac{2 m_{{\rm e}}(E-V)}{\hbar^2} \psi = 0
             \end{equation} 
         so that (8.12) is recovered. The time dependence of $\psi$ is given explicitly in~\cite{SchrodWM2} 
         as $\exp(2 \pi i \nu t)$, i.e.
         as a complex exponential, but this is not necessary for the above derivation 
         since (8.16) holds  also if $\psi$ is a real harmonic function (sine or cosine) of $t$.
         It may be remarked that the quantity $E$ in the Planck-Einstein relation is the relativistic
        energy of the electron whereas Eq.~(6.26) contains the non-relativistic bound state energy (c.f. Eq.(7.21), $E_{NR}$,
        (denoted by Schr\"{o}dinger as $E$) of Eq.~(6.25).
        Use of the
        Planck-Einstein relation here
       then implies the introduction of an unphysical negative frequency. This, together, with the unphysical
      nature of the phase velocity, $v_W(NR)$, introduced in (8.15) implies that this second `derivation' of the quantum
         wave equation should also be considered more as a heuristic exercise than a mathematically sound proof.  
         \par Neither of the above derivations of (7.27) requires the introduction into
        the equations of $\sqrt{-1}$. The spatial eigenfunctions that are solutions of (7.27) are real, not complex,
         functions of the spatial coordinates of the electron. 
        \par Schr\"{o}dinger finally derived the time-dependent quantum wave equation (4.7), after considerable
         discussion, in the first section of his fifth and last paper written in 1926 on wave
         mechanics~\cite{SchrodWM5}. This was done by introducing the ansatz that the time dependence of
          $\psi$ {\it is} given by a complex exponential:
           \begin{equation}
           \psi = u(x) \exp \left[\pm\frac{i E t}{\hbar}\right]
          \end{equation}
         so that 
             \begin{equation}  
              i \hbar\frac{\partial  \psi}{\partial t} = \mp E\psi 
             \end{equation}
              which when substituted in the last member of the last equation in (8.17) and transposing 
               gives:
              \begin{equation}  
              -\frac{\hbar^2}{2 m_{{\rm e}}}\nabla^2 \psi +V\psi \mp i \hbar \frac{\partial  \psi}{\partial t} = 0.
             \end{equation}
             On choosing the minus sign in (8.18) and (8.20) the time-dependent equation
             (4.8) is obtained. Here $\sqrt{-1}$ makes an essential and necessary appearence.
              However, Schr\"{o}dinger's ansatz (8.18) is just {\it the fundamental second postulate
             of Feynman's formulation of quantum mechanics for the time dependence of a path amplitude}. c.f.
               Postulate II, Eq.~(3.3) and Eq.~(6.18).
               It was therefore, in a sense, not necessary for Feynman to demonstrate that the 
              Schr\"{o}dinger equation (4.8) follows from his postulates, since, in fact, Schr\"{o}dinger
             had made, {\it a priori}, essentially the same postulate, (8.18), in order to derive the equation!

             \SECTION{\bf{Summary and discussion}}
              The H-J equation (2.6) is a first order partial differential equation for Hamilton's principal
              function $S = F(q_i,P_i,t)$, the generating function that transforms canonical coordinates and momenta,
              that respect Hamilton's equations, from $q_i$, $p_i$ to  $Q_i$, $P_i$, in such a way
              that the transformed Hamiltonian vanishes. The same function $S = \int L dt$,
              where $L$ is the Lagrangian of the dynamical system considered, occurs in the fundamental formula (3.3) 
              of Feynman's path integral formulation of quantum mechanics~\cite{FeynRMP}:
               $K_{{\rm BA}}^{(j)} = A_{{\rm BA}}^{(j)}\exp[i S_{{\rm BA}}^{(j)}/\hbar]$ giving the
               amplitude $K_{{\rm BA}}^{(j)}$ to find a particle,
               originally at space-time point A, at the point B, having followed the space time path $j$. 
              Writing $\psi \equiv K/A$ and substituting for $S$ in the H-J equation (4.7) for a 
              non-relativistic particle confined by a time-independent potential $V$ gives immediately the time-dependent 
               Schr\"{o}dinger equation (4.8) on identifyimg $\psi$ with the quantum mechanical
              wavefunction. Application of the Schr\"{o}dinger equation to the hydrogen atom then shows 
              that its spatial wavefunction, according to Feynman's first postulate, that combines
              quantum mechanical superposition with Born's probabilistic interpretation of quantum
               mechanics~\cite{BornPI}, is the probability amplitude to find the bound electron 
              at a particular spatial position in a particular bound state of the atom.
               This is in accordance with Dirac's more
              formal interpretation of the wavefunction as a transformation
              matrix element between representations with different eigenstates~\cite{DiracTFQM}.
                \par For a free particle, with relativistic energy $E$ and momentum $p$, the Hamilton's principal
              function takes the Lorentz-invariant form: ${\cal S} = px-Et$, as follows by integrating the 
             transformation equations in (2.4) and (2.5). The replacements $p/E \rightarrow v/c^2$ and $ x \rightarrow vt$ in
            this formula for ${\cal S}$ gives the relativistic Lagrangian:  ${\cal L} = -mc^2\sqrt{1-\beta^2}$ for a 
            free particle as well as the fundamental formulae of relativistic kinematics and space-time geometry in (5.10).
            The Planck-Einstein relation $E = h \nu$, (6.3) and the de Broglie relation  $\lambda = h/p$, (6.4)  are derived
            by mathematical substitution in the formula, (6.2), for the phase of the path amplitude for a free particle.
           . 
  \par The derivation by de Broglie of his wavelength relation by Lorentz transformation  of the path amplitute phase
         and use of the Planck-Einstein relation is compared with that obtained from Eq.~(6.2). Also de Broglie's
         derivation of a `group velocity' equal to the particle velocity $v$, by introducing the concept of a
         refractive index for `matter waves' in free space is compared with that obtained from the energy-momentum-mass
        relation (6.6) in conjunction with the Planck-Einstein and de Broglie relations. 
           \par Hamilton's discussion of the analogy between particle kinematics and geometrical optics, based on
              `wave-fronts' associated with Hamilton's characteristic function, $W$, in Eq.~(6.18) is used to derive an associated
             phase velocity: $v_W = E/|\vec{p}|$. For relativistic motion, where $E/|\vec{p}| = v/c^2$, 
              $v_W$ is equal to $v_{\phi}$. However for non-relativistic motion it is found that $v_W = v/2$ whereas 
             the relativistic formula $v_W = v_{\phi} = c^2/v$ must hold for all values of $v$. This inconsistency
            is indicative of the unphysical nature of the `matter waves' in free space and their associated `phase
            velocity'. This conclusion is reinforced by the gauge dependence of the wavefront direction,  phase velocity
            and wavelength of the `matter waves' associated with particle motion in magnetic fields~\cite{ES},
              to be contrasted with the gauge independence, and hence physical uniqueness,
      of the phase difference between the amplitudes of
              different paths.
             \par The path amplitude for a free particle is an eigenfunction of the differential operators 
             ${\cal P} \equiv -i\hbar\vec{\nabla}$,  ${\cal E} \equiv i\hbar \partial/\partial t$ with
               eigenvalues $\vec{p}$, $E$ respectively. Consideration of the application of these operators 
              to arbitary functions of space-time allows derivation of the canonical commutation relations
               (7.10) and (7.11), while their application to the wavefunction of an electron in a bound state of
               the hydrogen atom with the Hamiltonian of (7.26) enables derivation of the time-independent
                Schr\"{o}dinger equation for this wavefunction. It is argued that these path integral derivations
               have important pedagogical advantages over the presentation of the same subjects in Dirac's book on
                quantum mechanics~\cite{DiracFQC}.
   \par Schr\"{o}dinger's first published derivation of the time-independent quantum wave equation~\cite{SchrodWM1}
        also took as its primary postulate the H-J equation (4.7) for a non-relativistic particle constrained by
         a time-independent potential V. The relation between the wavefunction $\psi$ (introduced here for the
          first time) and Hamilton's principle function $S$, is given by 
       the ansatz: $S = K\ln\psi$ where $K$ is real constant with the dimensions of action, instead of the
       path-integral postulate $S = -i\hbar\psi$.The correct equation is obtained by the artifice of introducing the
       quantity $J$ defined in Eq.~(8.7) and requiring that
       it is stationary for arbitary variations of $\psi$.
        In fact, as pointed out by Heisenberg~\cite{HeisLagr}, the quantum wave 
       equation can be derived by requiring stationarity with respect to  variations of $\psi$ of the Lagrangian
       given in (8.13), i.e. by
       an application of Hamilton's principle, which is also the basis for Hamilton's equations and the H-J equation itself.
       \par  Schr\"{o}dinger's second published derivation of the time-independent quantum wave equation~\cite{SchrodWM2}
        was based on Hamilton's mechanical/geometric optics analogy discussed in Section 6. The equation is given on
         subsituting the non-relativistic phase velocity $v_W({\rm NR})$ of Eq.~(6.26) in the classical wave equation
        (8.15). Other postulates are a harmonic time dependence of $\psi$, and the 
        Planck-Einstein relation $E = h\nu$. This derivation, unlike that presented in Section 7 leading to (7.27) 
        requires the introduction, in an intermediate step, of the unphysical quantity $v_W({\rm NR})$.
        Also the energy in the Planck-Einstein relation is the relativistic energy given in Eq.~(5.10), whereas
       in the derivation it is equated with the non-relativistic energy $E_{NR}$ of Eq.~(6.26). Furthermore,
          the symbol $E$ in the equation is interpreted by Schr\"{o}dinger as the (negative) bound state 
            energy ($E^{{\rm BS}}_{{\rm H},i}$ of Eq.~(7.21)) of the hydrogen atom in order
       to obtain the atomic wavefunctions. This second derivation is, therefore, also mathematically flawed.
      \par To derive the time-dependent equation~\cite{SchrodWM5} Schr\"{o}dinger makes the ansatz:
      \newline $\psi = u(x)[\exp(\pm i E t/\hbar)]$,
        which, when used to eliminate $E$ from the previously derived time-independent equation, yields (4.8) on choosing
        the minus sign in the complex exponential. The ansatz used (except that a sign ambigity devoid of any 
       physical significance is retained) is exactly the time dependence of the path amplitude phase (c.f. Eq.~(6.18))
       specified by Feynman's postulate II, in the case of a time-independent Hamiltonian. It is then not surprising
       that Feynman was able to derive the Schr\"{o}dinger equation from his postulates!
       \par The work presented in the present paper is in some ways complementary to, and in others overlaps with,
        that presented in a previous paper by the present author~\cite{JHFEJP} published in this journal. In the latter
     several of the results of Section 7 of the present paper are derived starting from different considerations
         ---the `inverse correspondence', low photon density, limit of classical electrodynamics--- where quantum
            mechanical concepts become applicable.
         The basic premise of Ref.~\cite{JHFEJP} ---interpretation of the equations of classical electrodynamics
         as a description of photonic behaviour, is the same as that of the de Broglie paper~\cite{dBPM} of which
         the present author was unaware at the time of writing Ref.~\cite{JHFEJP}. As previously mentioned, the
         essential concepts of Feynman's formulation of quantum mechanics are also to be found in Ref.~\cite{dBPM}. 
   \par In all applications of the Feynman path amplitude formulation of quantum mechanics to actual 
        experiments care must be taken in the identification of all parameters relevant to the
        space time description. Some examples are described in Ref.~\cite{JHFAP}, For example, in 
         physical optics experiments where interference effects involving real photons are observed,
         then since in (6.2) $x/t = c = E/p$ it follows that $\phi  = (xp-Et)/\hbar = 0$, so, as pointed out
         by Feynman~\cite{FeynQED1}\footnote{~`Once a photon has been emitted there is no further turning
        of the arrow as the photon goes from one place to another in space-time.' i.e. there is no change in the
        phase of the photon path amplitude.}, the phase of the propagator of the photon vanishes! Detailed 
         calculation~\cite{JHFAP} shows that, in fact, the `photon wavefunction' actually contains the 
         phase of the decay amplitude of the photon's unstable source
        atom.

  \par {\bf Acknowledgement}
      I would like to thank an anonymous referee for a careful
      reading of an earlier version of this article and for many
      helpful suggestions that have improved both its content and clarity of presentation.
  
\end{document}